# Investigation on the Atmospheric Incoming Flow of a Utility-Scale Wind Turbine using Super-large-scale Particle Image Velocimetry

Cheng Li[1], Aliza Abraham[1,2], Biao Li[1], and Jiarong Hong[1,2]

[1]Department of Mechanical Engineering, University of Minnesota, Minneapolis, MN 55455, USA
[2]St. Anthony Falls Laboratory, University of Minnesota, Minneapolis, MN 55414, USA

**Abstract:**

The atmospheric incoming flow of a wind turbine is intimately connected to its power production as well as its structural stability. The overall relationship is traditionally expressed in terms of a power curve acquired using 10-min mean of hub-height wind speed and power output without considering the details of the incoming flow field. Very little field data has been collected investigating instantaneous behaviour of wind turbine interacting with incoming flow. In this paper, we present a whole-field incoming flow measurement of a 2.5 MW pitch-regulated turbine at high spatio-temporal resolution, using super-large-scale particle image velocimetry (SLPIV) with natural snowflakes. The datasets include over one hour duration of incoming flow with an effective field of view of 85 m (vertical) × 40 m (streamwise) centered at 0.2 rotor diameter upstream of the turbine. The mean flow field shows the presence of the induction zone and a distinct region with enhanced vertical velocity. In comparison to predictions from vortex theory, SLPIV streamwise velocity at hub height shows a steeper velocity drop close to the rotor plane and a more confined induction effect with the velocity drop becoming negligible at around 0.4 rotor diameters upstream. Time series of nacelle sonic anemometer and SLPIV measured streamwise velocity outside the induction zone show generally matched trends with time-varying discrepancies potentially due to the induction effect and the flow acceleration around the nacelle. These discrepancies between the two signals, characterized by the sonic-SLPIV velocity ratio, is normally distributed and is less than unity 85% of the time. The velocity ratio first decreases with increasing incoming wind speed up to around the rated speed of the turbine, then plateaus, and finally rises with further increase of wind speed. With conditional sampling, the distribution of the velocity ratio shows that larger yaw error leads to an increase in both the mean and the spread of the distribution. Moreover, as the incident angle of the incoming flow changes from negative to positive (i.e. from pointing downward to upward), the velocity ratio first decreases as the angle approaches zero. With further increase of the incidence angle, the ratio then plateaus and fluctuations are augmented. Finally, our results show that the intensity of short-term velocity fluctuation has limited impact on the sonic-SLPIV velocity ratio.

# 1. Introduction

Understanding the characteristics of the atmospheric turbulent flow approaching the wind turbine (referred to as the incoming flow hereafter) plays a crucial role in improving the turbine operation for better energy extraction efficiency and structural reliability. Specifically, for example, the relationship between energy extraction and incoming flow is traditionally characterized by the power curve without considering the details of the incoming flow. Informative as it is, a significant amount of uncertainties have been observed under various applications using the power curve [1,2]. The uncertainties have been demonstrated to be associated with the incoming flow conditions, such as wind shear, inflow angle, wind direction profile, turbulence intensity, turbulent kinetic energy, and researchers have been developing various methods to accurately model the power curve [3–7]. Furthermore, utilizing detailed incoming flow information, researchers have proposed various control algorithms ranging from simple pitch control to so-called 'smart rotor control' in an aim to retain high energy output and structural reliability [8–12]. From a wind farm perspective, the wake development of a wind turbine is also influenced by the atmospheric conditions and turbulence intensity of the incoming flow, which is important for the overall optimization of wind farm operation [13].

Despite a significant amount of work focusing on the turbine wakes [14–22], however, only a limited number of studies have been dedicated to the incoming flow. Generally, as the flow approaches a wind turbine, it slows down due to the turbine blockage or the induction effect, and the region of this flow deceleration is referred to as the induction zone [23]. In theory, the induction zone is traditionally analyzed by considering the turbine rotor as an actuator disc under steady flow conditions [24,25], which allows pressure difference across the rotor plane without a velocity discontinuity. Accordingly, the effect of the induction zone can be characterized using the axial induction factor $a = (U_\infty - U_R)/U_\infty$, where $U_\infty$ and $U_R$ refer to the freestream and rotor plane velocity, respectively. Furthermore, the velocity distribution along the center of the rotor upstream could be analytically expressed by applying the Biot-Savart law [23,26]. Although these theoretical analyses can provide meaningful insight and general characterization of the induction effect, they are only valid for mean flow and the dynamic behavior of the induction zone, especially under complicated atmospheric turbulence, remains largely unknown. Such information is of great practical relevance for the improvement of turbine controls since most control algorithms employed in utility-scale turbines still rely on simply the feedback control of the mechanical systems [27], without utilizing the knowledge of the actual temporal variation of the incoming flow.

The laboratory studies on the incoming flow are generally conducted using scaled down turbine models, with Reynolds number, $Re = U_\infty D/v$, ranging from $10^4 \sim 10^5$, where $D$ is the rotor diameter, $v$ is the kinematic viscosity of air. Using both particle image velocimetry (PIV) and hot-wire experiments, Medici *et al.* quantified the induction effect of several model wind turbines under a range of flow speeds [23]. The induction effect is found to be discernable up to three times the rotor diameter, and the upstream velocity deficit is found to be consistently larger than that predicted by vortex sheet theory. Howard and Guala conducted wind tunnel PIV measurements of the incoming flow focusing on temporal incoming flow and turbine performance correlations [27]. By using power-velocity correlations, they showed the wind speed at about $D/4$ above hub height is most correlated with power, consistent with previous studies on the rotor equivalent wind speed defined based on kinetic energy flux over the swept rotor area for power curve measurement [28]. With both PIV and hotwire measurements, Bastankhah and Porté-Agel investigated the incoming flow characteristics under both regular and yawed conditions, and showed that the induction zone becomes asymmetric under yawed conditions associated with increasing nacelle frontal area and velocity asymmetry relative to the rotor plane [29]. Moreover, they showed the existence of very long structures in the induction zone with scales comparable to the size of the rotor. However, due to the significant difference between the lab and the field conditions, the community still relies on the field campaigns under varying atmospheric conditions to provide relevant information about the incoming flow for utility-scale turbine operations [23,27,30].

Using three LiDAR scan patterns, Simely *et al.* investigated the region up to $1.5D$ upstream of a 225 kW wind turbine at various incoming flow speeds. The results showed that the measured velocity reduction increases to 7.4% at the rotor tip region and 18% near the hub when the turbine operates at high power, and decreases to 6% near the hub for low power cases. They also pointed out that the standard deviation of the streamwise velocity is suppressed approaching the rotor plane whereas that for the vertical velocity is enhanced. Howard and Guala deployed LiDAR in front of a 2.5 MW wind turbine at multiple locations upstream to measure the vertical velocity profile, showing a cross-correlation between incoming flow and turbine power similar to their wind tunnel study [27]. However, limited by spatial and temporal resolution of the field measurement techniques, the temporal variations of the incoming flow at different elevations within the induction zone and their correlation with turbine operation were not investigated in these field studies.

The induction zone has been studied numerically through a series of work by Frosting using Reynolds-averaged Navier Stokes and detached-eddy simulation [31]. They showed that thrust is the major parameter governing the induction strength [32], and other turbine parameters, such as yaw, tilt, and rotor design, etc. [32–34]. and environmental characteristics including wind shear, atmospheric turbulence, and topography, etc. [35] have limited impact on the induction zone. As a result, based on numerical simulations, a simple model has been proposed to relate rotor centerline velocity to thrust coefficient, freestream velocity, and upstream locations [33]. Nonetheless, these numerical studies only applied actuator discs or lines models, without resolving turbine geometry, thus are not able to provide sufficient information on the flow around the nacelle.

Such information is particularly important for the operation of current utility-scale wind turbines, which are equipped with a nacelle mounted anemometer for yaw control, turbine start-up and shutdown. However, it is well-known that nacelle measurements involve substantial amount of uncertainties due to the induction effect and the unsteady flow around nacelle, as demonstrated by a series of prior studies. Specifically, by conditional sampling of field data, Antoniou and Pedersen found that pitch, anemometer location, yaw error, and terrain all contribute to the uncertainties and they proposed, assuming these conditions unchanged, the relation between the incoming flow and nacelle measurements could be curve-fitted well to a $5^{th}$ order polynomial [36]. This empirical and wind turbine specific relation, known as the Nacelle Transfer Function (NTF), is crucial for turbine power performance characterization [37]. Diznabi investigated the effects of nacelle anemometer locations on NTF [38]. In contrast to Antoniou and Pedersen [36], results show NTFs are almost linear regardless of location, but with decreasing slopes, approaching 1 away from the rotor. After phase averaging, all the velocity components exhibit signatures of blades passing with its intensity decreasing away from the rotor plane. Martin *et al.*[39] reported different atmospheric conditions (e.g. stratification and turbulence intensity, etc.) can result in different NTFs for the same turbine. Several numerical studies have also been conducted to obtain detailed flow around nacelle for predicting NTFs under different turbine conditions [40–42]. Notably, based on the experimental settings of Diznabi [38], Zahle and Sørensen [41] conducted three-dimensional rotor and nacelle resolved RANS simulations, and showed the flow field around the nacelle is highly unsteady casuing the wind measurement at the nacelle to be very sensitive to the location of the sonic anemometer. The study showed that, due to dispacement of root vortices, both yaw and tilt angles of the rotor plane can significantly modify profiles of wind speed, with nacelle velocity peaks shifting upward and downward for increasing tilt and yaw, repectively. Note that most these studies are based on long term (minutes to hours) averages, and the instantaneous response to these above-mentioned factors affecting the discrepancies between incoming flow and nacelle is mostly unknown.

In this current study, with the implementation of super-large-scale particle image velocimetry (SLPIV) [43–45], we are able to perform a detailed characterization of the incoming flow approaching a 2.5 MW turbine with high spatial and temporal resolution. Such information is used to investigate the induction zone and to assess the discrepancy between instantaneous incoming flow and the corresponding nacelle sonic anemometer measurements under different inflow conditions. The paper is structured as follows: section 2 provides a description of our experimental methods followed by the results in section 3, and the summary and the discussion of the results are provided in section 4.

## 2. Experimental Methods

### 2.1 Field site

The measurements of the incoming flow were conducted in the University of Minnesota EOLOS Wind Research Field Station in Rosemount, Minnesota. The station consists of a 2.5 MW three-bladed, horizontal-axis Clipper Liberty wind turbine (referred to as the EOLOS turbine hereafter) with pitch-regulating capability and a 130 m tall meteorological tower located 170 m south of the turbine.

The met tower is equipped with wind velocity wind velocity (sonic, and cup-and-vane anemometers), pressure, humidity and temperature sensors installed at elevations ranging from 7 m to the highest point of the rotor, i.e. 129 m. High temporal resolution (20 Hz) Campbell Scientific CSAT3 3D sonic anemometers are installed at 10 m, 30 m, 80 m, and 129 m. These four heights are intended to match the rotor's top, hub, and bottom as well as standard 10 m height. Wind speeds at elevations 3 m below each of the CSAT3's (elevations of 126, 77, 27 and 7 m) and at elevations representing midpoint between the edge of the rotor and hub height (105 and 55 m) are measured using standard 1 Hz cup-and-vane anemometers.

The turbine has a rotor diameter ($D$) of 96 m and a support tower of 80 m in height ($H_{\text{hub}}$), and attains the rated power at a velocity ($U_{\text{rated}}$) of 11 m/s. As shown in Figure 1, the nacelle is slightly tilted 5.5° above horizontal and the blades are pre-coned with a 3.3° angle to achieve blade-tower clearance. A sonic anemometer is installed at around 1.5 m above the nacelle ceiling (corresponding to an elevation of 84 m above the ground) to measure wind speed ($u_{\text{N}}$) and direction ($\varphi_{\text{N}}$) with an effective sampling rate of 1 Hz. Additionally, other turbine operational parameters, including turbine power ($p$), blade pitch ($\beta$), and yaw angle ($\gamma$) can be obtained from the turbine SCADA database at a sampling rate of 20 Hz. In the past few years, the site has been used for a number of field campaigns for the study of wind turbine wakes [43,45], atmospheric boundary layer [44,46], and snow settling dynamics [47]. The details on the facility of the EOLOS station can be found in the supplemental material of Hong et al.[43]

## 2.2. Super-large-scale particle image velocimetry (SLPIV) measurements

The experiments employ the SLPIV technique from Dasari et al.[45] to investigate the incoming flow field of the EOLOS turbine. This technique utilizes the natural snowfall as flow tracers introduced in Toloui et. al.[44] and was validated through a series of follow-up studies recently [45–47]. Briefly, the experimental setup is composed of an optical assembly for illumination, a camera and the corresponding data acquisition system. The optical assembly includes a 5 kW highly collimated search light and a curved reflecting mirror for projecting a horizontal cylindrical beam into a vertical light sheet. The camera, Sony-A7RII equipped with a 50 mm $f/1.2$ lens and operating at 60 Hz, is used to provide full HD resolution videos, corresponding to a sensor size of 1920 × 1080 pixels, for SLPIV measurement and flow visualizations.

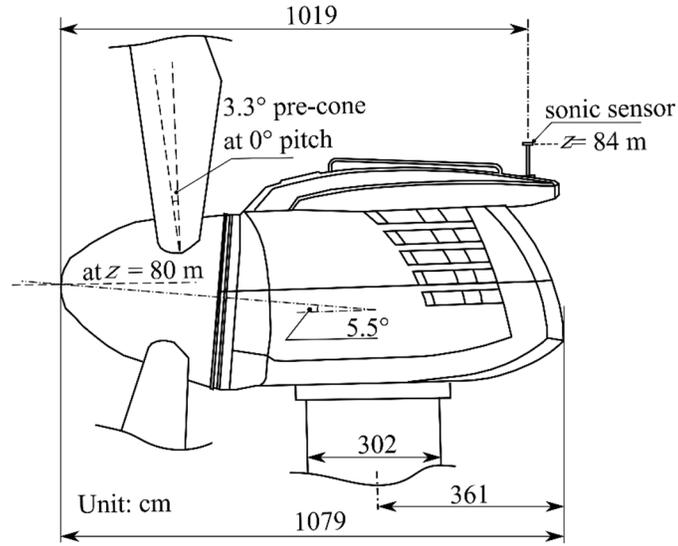

Figure 1. Schematic showing the dimension and detailed geometry of the nacelle of the EOLOS turbine.

| | Start time (CST) | Duration | $U_N \pm u_{N,std}$ (ms$^{-1}$) | $U_{met} \pm u_{met,std}$ (ms$^{-1}$) | $U_{0.4D} \pm u_{0.4D,std}$ (ms$^{-1}$) | $\Phi_N \pm \varphi_{N,std}$ | RH % | TI % | Thermal Stability | |
|---|---|---|---|---|---|---|---|---|---|---|
| | | | | | | | | | $\zeta = z/L$ | $Ri_B$ |
| **Run 1** | 21:30:21 | 24 min 58 s | 12.52 ± 2.17 | 13.53 ± 2.32 | 13.44 ± 2.27 | 287.5° ± 7.64° | 93.6 | 19.1 | -0.007 | -0.05 |
| **Run 2** | 22:02:08 | 19 min 40 s | 12.71 ± 2.53 | 14.10 ± 2.36 | 13.56 ± 2.20 | 285.8° ± 8.79° | 92.6 | 17.5 | -0.005 | -0.05 |
| **Run 3** | 22:23:07 | 19 min 18 s | 13.65 ± 2.60 | 15.95 ± 2.78 | 14.53 ± 2.20 | 287.6° ± 8.34° | 92.1 | 18.6 | -0.006 | -0.04 |

Table 1. Overview of SLPIV Run 1 – 3. $U_N \pm u_{N,std}$ – nacelle sonic measured mean ± standard deviation of wind speed, $U_{met} \pm u_{met,std}$ – met tower sonic measured ($z = 80$ m) mean ± standard deviation of wind speed, $U_{0.4D} \pm u_{0.4D,std}$ – SLPIV measured mean ± standard deviation of wind speed at $x/D \simeq -0.4$, $\Phi_N \pm \varphi_{N,std}$ – nacelle measured mean ± standard deviation of wind direction, $RH$ – relative humidity, $TI$ – turbulence intensity, $\zeta$ – Monin-Obukhov ratio, $Ri_B$ – bulk Richardson number.

The data for the present study is obtained during a field campaign from 21:00 to 23:00 on December 4$^{th}$, 2017. The details regarding the dataset are summarized in Table 1. The methods for calculating the corresponding meteorological conditions are discussed in section 2.3. Figure 2 illustrates the experimental setup for the deployment. Figure 2(a) shows a Google map of the deployment. The light sheet was deployed upstream of the turbine, with the light sheet aligning with the predominant wind direction 30 min before the experiment. Figure 2b illustrates the general coordinate system used for the present study with the origin at the hub front projection on the ground, the $x$ axis pointing downstream of the incoming flow, $y$ axis perpendicular to light sheet, and $z$ axis upward. Note, according to this definition, the hub front is a $x = 0$, $z = H_{hub} = 80$ m, and the sonic measurement at turbine nacelle is located at $x = 10$ m, $z = 84$ m. With a camera tilt angle $\theta = 21.1°$ and camera-light distance $L_{CL} = 188$ m, the center of the field of view is at $(x_{FOV}, y_{FOV}, z_{FOV}) = $ (-23 m, 0 m, 74 m) with a dimension of $H_{FOV} \times W_{FOV} = 135 \times 81$ m$^2$.

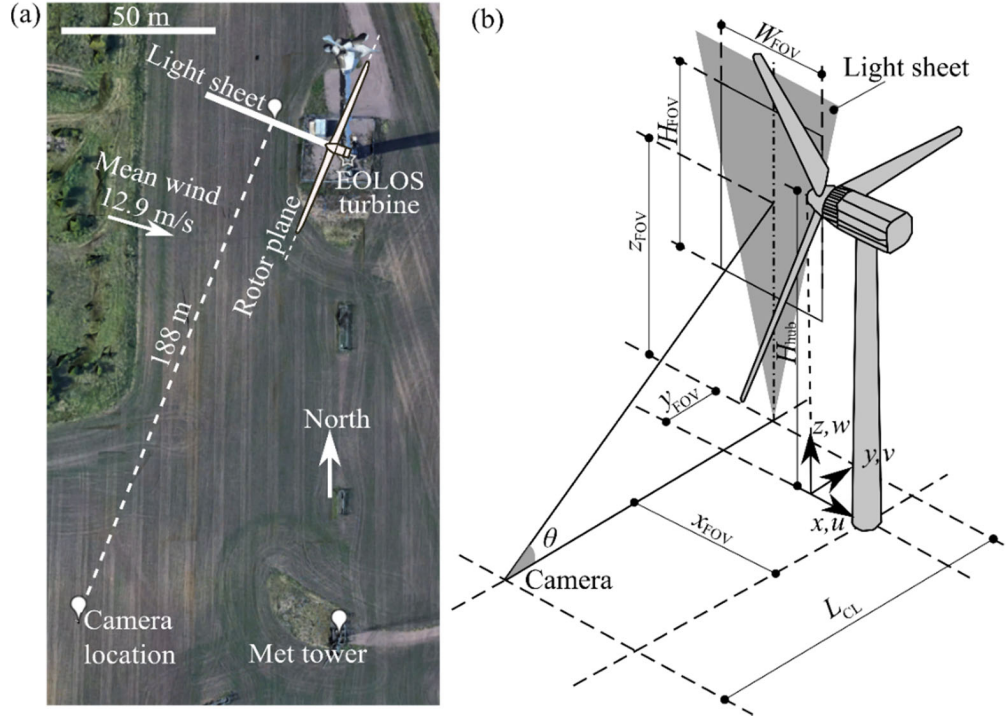

Figure 2. (a) A Google map showing the location of the wind turbine, met tower, and SLPIV experimental setup during the deployment on December 4$^{th}$, 2017. (b) Schematic of the measurement setup used in the deployment.

Note that the conditions related to the experimental setup orientation (i.e. $\varphi_N$, $\gamma$) change constantly during the deployment. Figure 3(a) provides a time series of these parameters and light sheet orientation with the shaded areas indicating periods of Runs 1 - 3, consecutively. The measured wind direction during Runs 1 - 3 ranges from 268° to 310° with an arithmetic mean of 287° aligning almost perfectly with the light sheet direction (289°). The corresponding yaw error ($\Delta\gamma = \varphi_N - \gamma$) quantifying the alignment of the rotor plane to the wind direction ranges from -20° to 26°. Moreover, except for three instances with relatively large yaw angle variations, the yaw angle remains in the 286 - 290° range, resulting in a good alignment of light sheet to be perpendicular to the rotor plane. During the three runs, the wind speed was fluctuating around $U_{rated}$ as shown in Figure 3(b), with a mean of 12.9 m/s, and the turbine was operating at control regions of 2.5 and

3. The turbine operational region is based on the rotor speed and power constraints of the turbine and is determined by incoming flow and the control strategy. Power output stays at the rated power at Region 3 with increasing wind speed. In order to mitigate strong wind, pitch angle increases almost linearly with increasing wind speed for winds over $U_\text{rated}$.

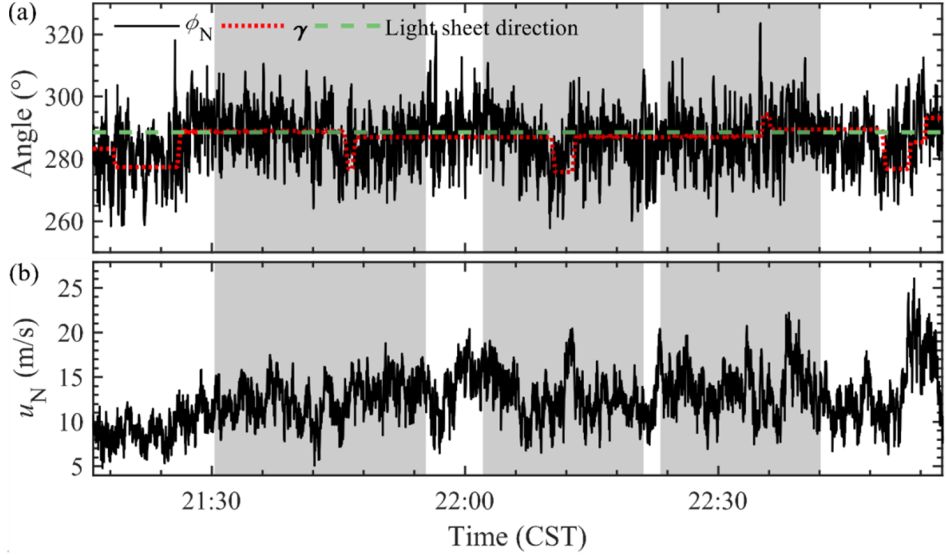

Figure 3. Time series of (a) wind direction superimposed with the nacelle, and light sheet direction and (b) wind speed for the time duration of the deployment on December 4$^\text{th}$, 2017. Note that the gray vertical bands mark the SLPIV video data collection periods.

## 2.3. Metrological condition

For each run, SCADA sonic and met tower measurements are used to quantify the overall meteorological conditions, with data provided in Table 1. Both mean and standard deviation of nacelle wind speed ($U_\text{N} \pm u_\text{N,std}$) and direction ($\Phi_\text{N} \pm \varphi_\text{N,std}$) are calculated using nacelle sonic measurements. Note, as will be shown later in section 3, such values will be impacted by the presence of wind turbine, and thus different from the freestream value for the wind turbine. As a freestream reference, also provided are the met tower sonic wind speed ($U_\text{met} \pm u_\text{met,std}$) at $z = 80$ m. The lateral distance between the EOLOS turbine and met tower according to the mean wind direction is around 163 m. Also provided are SLPIV measured incoming flow velocities ($U_\text{0.4D} \pm u_\text{0.4D,std}$), evaluated at $x/D \simeq -0.4$ upstream (i.e. close to the most upstream position of SLPIV measurements.

The thermal stability is quantified using both Moin-Obukhov ratio and bulk Richardson number in current study. The Moin-Obukhov ratio, $\zeta = z/L$, is calculated using met tower height at $z = 10$ m and the Moin-Obukhov length, $L = -u_*^3 \overline{\theta_v}/\kappa g \overline{w'\theta_v'}$, where $u_*$ is the friction velocity, $\overline{\theta_v}$ is the average virtual potential temperature, $\kappa$ is the von Karman constant, $g$ is the gravitational constant, and $\overline{w'\theta_v'}$ is the virtual potential heat flux at the surface. The friction velocity is estimated using $u_*^2 \approx \overline{u'w'}$, and $\overline{w'\theta_v'}$ was approximated using turbulent heat flux, $\overline{w'T'}$. All these quantities are calculated from data acquired at $z = 10$ m. The bulk Richardson number, $Ri_\text{B} = g\Delta T \Delta z / T_0 \Delta U_\text{h}^2$, evaluated between the lowest ($z = 10$ m) and the highest ($z = 129$ m) sensors of the met tower, where $T_0$ is the mean temperature in Kelvin, $\Delta T = T(z = 129 \text{ m}) - T(z = 10 \text{ m})$ is the temperature difference, $\Delta z = 119$ m, $\Delta U_\text{h} = U(z = 129 \text{ m}) - U(z = 10 \text{ m})$ is the streamwise wind velocity difference.

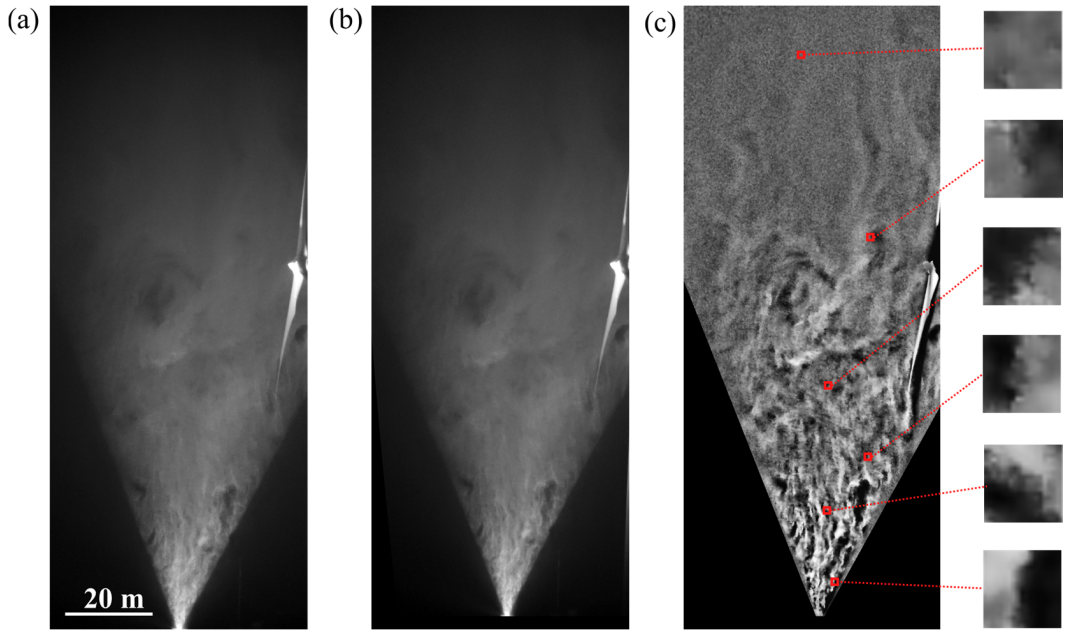

Figure 4. A sample of (a) raw image, (b) distortion corrected image, and (c) enhanced image after background subtraction, adaptive histogram equalization, and masking from run 1. Insets in (c) illustrate coherent structures at multiple elevations with the same size as the interrogation window (48 × 48 pixels) used in cross-correlation.

## 2.4. SLPIV data processing

To quantify the incoming flow velocity field, the SLPIV data processing generally follows procedures outlined in Dasari *et al.* [45], including distortion correction, image enhancement, as well as velocity field calculations using *LaVision Davis 8* software package by tracking the large-scale snow patterns and associated coherent structures. Figure 4 provides a sample image with raw, distortion corrected, and enhanced versions from Run 1. Note that the current spatial resolution (7.4 cm/pixel) is not capable of resolving individual snow particles like in cases with finer spatial resolution [43,44,46,47]. Thus, similar to Dasari *et al.* [45], the current velocity field analysis relies on tracking coherent structures within the atmospheric boundary layer. The insets in Figure 4(c) illustrate visible snow patterns at varying elevations in our FOV. The signal to noise ratio decreases with increasing elevations, especially after the hub height is reached.

The final interrogation window for the PIV cross-correlation is 48×48 pixels with 50% overlap, corresponding to a spatial resolution of 1.8 m/vector. To ensure proper displacement of the snow patterns, the cross-correlation is applied to the image pair with 5-frame skip, resulting in a temporal resolution of 12 Hz. Such spatial and temporal resolutions enable us to quantify the spatio-temporal evolution of the incoming flow field within the induction zone of the turbine and to assess the accuracy of the nacelle sonic measurements for charactering the incoming flow. To ensure the quality of velocity vectors from SLPIV, during post-processing, the vectors with correlation values less than 0.7 are rejected for the following analysis. As illustrated in the insets of Figure 3(c), due to the substantial decrease of signal to noise ratio of the image at higher elevations, the percentage of rejected vectors increases significantly away from the ground. Therefore, in the following analysis, the flow field is limited to locations where rejected vectors are less than 10% to avoid statistical bias. In addition, due to the settling of snowflakes, the velocity vectors within the flow field of SLPIV has an average (both spatially and temporally) downward vertical velocity of 1.6 m/s. This velocity is removed from the instantaneous velocity fields in accordance with the procedure followed by the prior studies. For further details regarding the SLPIV measurements using snowflakes, including the traceability of the snowflakes and the measurement uncertainty, please refer to our prior publications [44,45].

## 3. Results

### 3.1 Mean flow characteristics of incoming flow

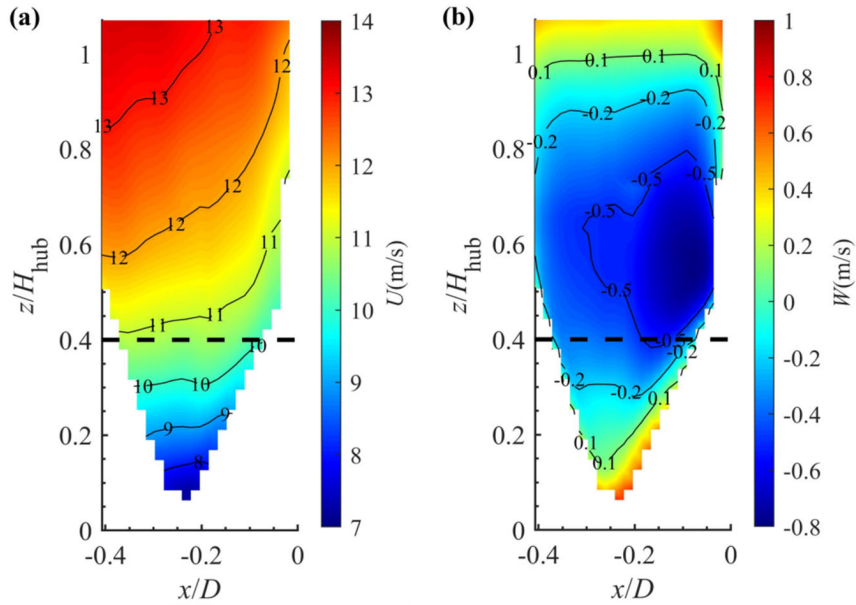

Figure 5. Time-averaged SLPIV velocity vector field of Run 1 (1:3 skip applied in both horizontal and vertical directions for clarity) superimposed with (a) mean horizontal velocity ($U$), and (b) mean vertical velocity ($W$) contours and iso-velocity lines from 8 to 13 m/s and -0.5 to 0.1 m/s, respectively. The black dashed line indicates the lowest level of the rotor plane, i.e. $z/H_{hub} = 0.4$.

Due to the slightly varying conditions (e.g. wind direction and turbulence intensity), only Run 1 data is included in this section to provide a statistically consistent analysis of the mean flow field approaching to the turbine. Figure 5a shows the time-averaged streamwise mean velocity ($U$) field. As described in the method section, the left ($x/D \simeq 0.4$) and upper boundary ($z/H_{hub} \sim 1.1$) of the field of view is determined by the percentage of the rejected vectors. The mean field clearly exhibits an induction region where $U$ decreases approaching the rotor plane at all specific elevations. Because of the turbine-induced blockage, the iso-velocity contours of the incoming flow near the ground are deflected upward approaching rotor plane. This pattern is similar to the results from a recent field measurements using vertical LiDAR scan around a smaller scale turbine with $D = 27$ m, $H_{hub} = 32.5$ m, lower wind speed $U_\infty = 7.0$ m/s, and higher induction factor $a = 0.25$ [30]. Moreover, both results show a much more confined region of induction compared to that predicted by vortex theory (i.e. the velocity profile reaches 5% reduction at $x/D \simeq -0.1$ in our study and $x/D \simeq -0.4$ for Simley *et al.* [30]; the locations predicted by vortex theory are around $-0.4$ and $-0.7$, respectively). The corresponding mean vertical velocity ($W$) is shown in

Figure 4b. As flow approaches the rotor plane, there is a distinct region with enhanced downward velocity centered at around $z = 45$ m or $(z - H_{hub})/D = -0.36$ and extending to the lower boundary of the rotor plane, similar to results from previous lab experiments[23,29], field LiDAR measurements [30], and numerical simulations [31]. For prior studies including lab, simulation and field LiDAR measurements, the vertical velocity reaches maximum at around the tip of the blades. In comparison, the current data shows the vertical velocity peaks at around $0.14D$ above bottom of the rotor plane. This discrepancy may be associated with the pre-cone angle of the EOLOS turbine rotor, blade deflection, and other geometric features of utility-scale turbines, which are not fully accounted for in the turbine models of the prior studies.

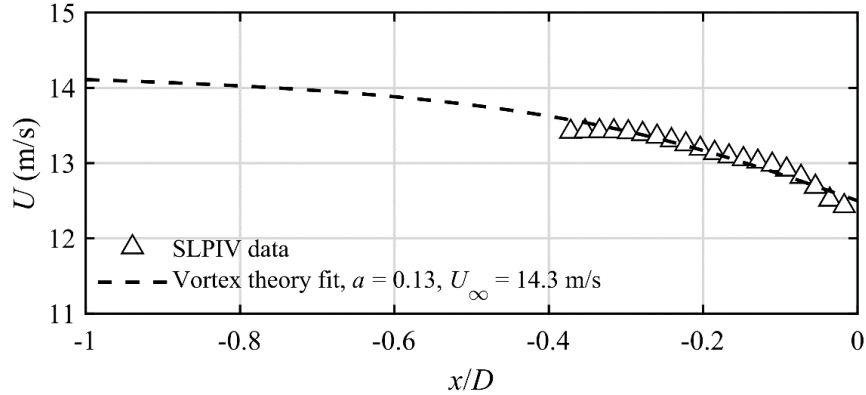

Figure 6. Mean streamwise velocity at hub height obtained from SLPIV data least square fitted with the analytical formula based on the vortex theory.

To obtain the induction factor ($a$) and the free stream wind speed approaching the turbine ($U_\infty$), the SLPIV data is fitted (least square sense) with the analytical formula proposed by Medici et al. [23] as shown in Figure 6. According to Medici et al.[23], the incoming flow speed along the turbine axis can be expressed as

$$\frac{U(x)}{U_\infty} = 1 - a\left[1 + \frac{\frac{2x}{D}}{\sqrt{1+\left(\frac{2x}{D}\right)^2}}\right] \quad (1)$$

The figure presents the SLPIV data spatially averaged around hub height ($z = 80$ m) with a vertical span of $\pm 1$ m. As the figure shows, the experimental data fits reasonably well in the majority of the streamwise measurement span, within a maximum deviation of 8% from the theoretical prediction of streamwise velocity reduction, resulting in an induction factor $a = 0.13$ and $U_\infty = 14.3$ m/s. Nevertheless, the SLPIV data presents a steeper velocity drop near the rotor plane (at $x/D \gtrsim -0.1$) than that predicted by the vortex theory. This may be caused by the additional induction effect associated with the presence of the rotor hub and the turbine nacelle that are not considered in the simplified vortex theory. Moreover, the SLPIV result starts to plateau at $x/D \simeq -0.3$ with further increase of upstream distance from the turbine. To evaluate the accuracy of the nacelle sonic in measuring the incoming flow in Section 3.2, we use the SLPIV data acquired at $x/D \simeq -0.4$ as a surrogate for the incoming flow velocity, focusing on the nearfield of the induction zone. Based on data provided in Table 1, met tower measured mean velocities are in slightly higher than SLPIV measurements, indicating SLPIV measurements at $x/D \simeq -0.4$ is an underestimation of the incoming flow, however the signal at $x/D \simeq -0.4$ remains highly correlated with nacelle sonic data, enabling us to compare both signals instantaneously. Note that as mentioned in the Introduction Section, in the past studies, incoming flow and nacelle measurements comparison are mostly based on a 10-min or longer averages [30,36,39].

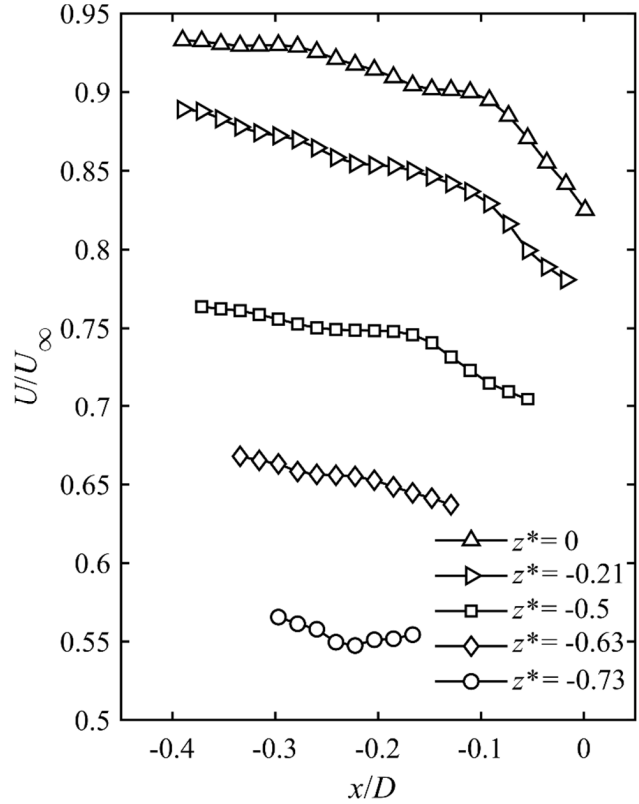

Figure 7. The variation of the SLPIV mean streamwise velocity approaching the turbine at multiple elevations. To accentuate the elevation with respect to the hub height, the nondimensionalized elevation $z^* = (z - H_{\text{hub}})/D$, is employed in the figure.

Figure 7 presents the variation of the SLPIV mean streamwise velocity approaching the turbine at multiple elevations relative to hub height. A nondimensionalized elevation, $z^* = (z - H_{\text{hub}})/D$, has been used to indicate the relative distance to the hub height. As it shows, the streamwise velocity decreases for all elevations approaching the rotor plane. However, the rate and the magnitude of velocity decrease differ at different elevations. The overall effect of induction is significantly stronger within the rotor plane ($|z^*| \leq 0.5$) than that outside the rotor plane ($|z^*| > 0.5$). Within the elevation span of the rotor, the velocity reduction of the incoming flow is 12% at hub height ($z^* = 0$) and decreases to 10% at $z^* = -0.5$. Outside the rotor span, Figure 7 shows around 7% velocity reduction at $z^* = -0.63$ and 5% at $z^* = -0.73$. These results qualitatively agree with those from the field measurements on a 225 kW turbine using vertical LiDAR scan [30]. Specifically, for data from Case 8, the velocity reduction is around 15% at hub height, and 10% at the edge. The larger vertical velocity reduction variation is likely associated with the differences in incoming flow profile as well as the turbine scale.

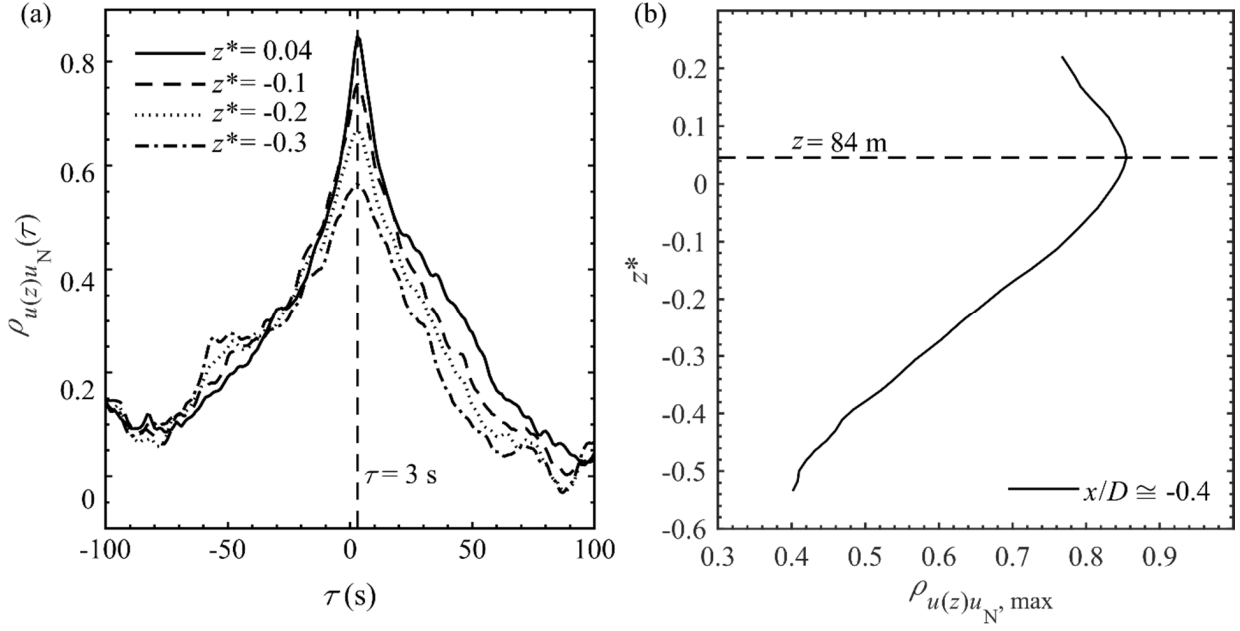

Figure 8. (a) Temporal correlation between SLPIV data at $x/D \simeq -0.4$ and the nacelle sonic data with different time lags at multiple elevations. (b) The profile of maximum temporal correlation coefficients of SLPIV and the nacelle sonic data.

## 3.2 Comparison between SLPIV and nacelle sonic measurements

SLPIV provides flow measurements with sufficient spatio-temporal resolution in the field very near the turbine upstream to evaluate the performance of the nacelle-mounted sonic anemometer as a common instrument to probe incoming flow for turbine operations. In this section, the SLPIV data from all three runs are employed, and the data at the very upstream of the SLPIV measurements (i.e. at $x/D \simeq -0.4$) is selected to represent the incoming flow. As the first step, we conduct the temporal correlation of the SLPIV data at each elevation with the nacelle sonic measurements, i.e. $\rho_{u(z)u_N}(\tau)$, to determine the elevation at which the SLPIV is most correlated with the nacelle sonic data for the follow-up comparison. For each elevation, the temporal correlation between the two measurements reaches a maximum ($\rho_{u(z)u_N,max}$) at a time lag of 3 s as shown in Figure 8a. Such time lag corresponds approximately to the advection time between the locations of the two measurements. Subsequently, $\rho_{u(z)u_N,max}$ is calculated for each elevation as shown in Figure 8b. The corresponding profile of $\rho_{u(z)u_N,max}$ peaks with a value of 0.85 at $z = 84$ m, which matches precisely to the elevation of nacelle-mounted sonic anemometer. The $\rho_{u(z)u_N,max}$ decreases significantly away from the peak with its value more than halved at the bottom of the rotor plane, i.e. at $z^* = -0.5$. Therefore, for comparison with the nacelle sonic measurements in the following analysis, we use the SLPIV data at the nacelle sonic elevation, denoted as $u_{0.4D}$.

Figure 9 shows a comparison of streamwise velocity between SLPIV at $x/D \simeq -0.4$ and the nacelle sonic measurements at the same elevation $z = 84$ m. The SLPIV data down-sampled to match the 1 Hz temporal resolution of the nacelle sonic for consistent comparison. The general trend of SLPIV measurements matches well with that of the nacelle sonic, indicating a good reliability of the measurements obtained from the two independent approaches. Nevertheless, the discrepancy between two measurements persists at all times with varying degrees. Specifically, the nacelle measurement ($u_N$) yields a mean velocity of 12.9 m/s, which is about 7% lower than the incoming flow measured from SLPIV ($u_{0.4D}$). As shown in Figure 9, for most of the time, $u_N$ shows a stronger fluctuation compared to that of $u_{0.4D}$. Such time varying discrepancies are likely to be a combined result of induction effect and the unsteady flow around the nacelle, which will be further investigated in the remaining part of this section.

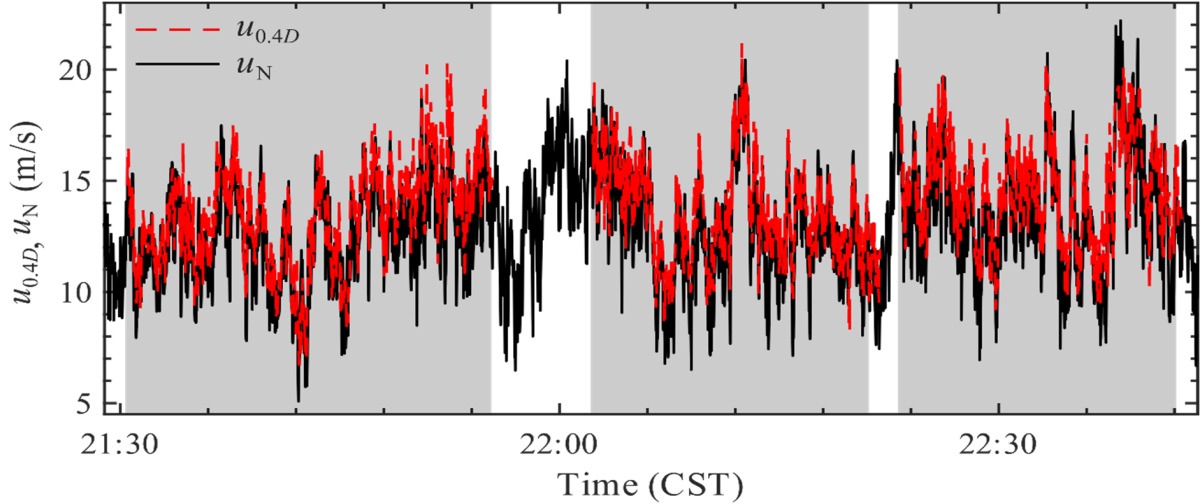

Figure 9. Comparison between SLPIV at $x/D \simeq -0.4$ and the nacelle sonic measurements at the same elevation $z = 84$ m. Note that the gray vertical bands mark the SLPIV video data collection periods.

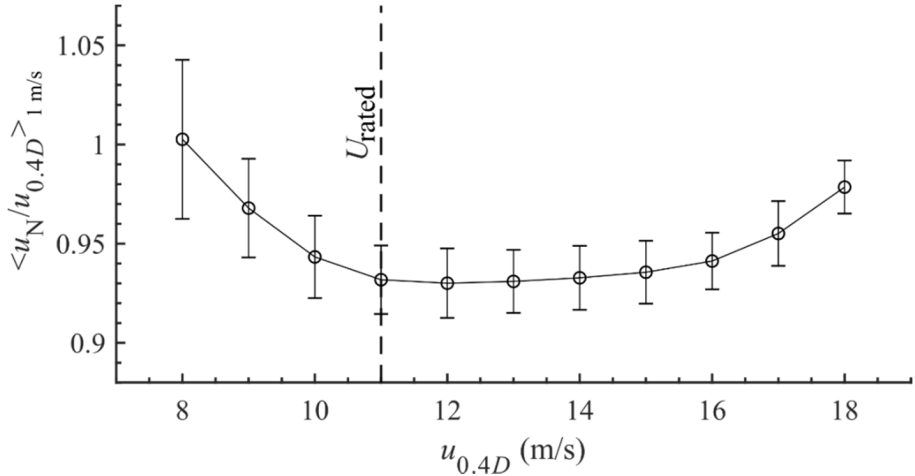

Figure 10. Effect of incoming wind speed on the sonic-SLPIV velocity ratio.

To characterize quantitatively the discrepancy between the nacelle sonic measurement and the incoming flow represented by SLPIV measurements at $x/D \simeq -0.4$, we introduce the sonic-SLPIV velocity ratio defined as $u_N/u_{0.4D}$. Figure 10 shows the ensemble-averaged sonic-SLPIV velocity ratio over Runs 1 – 3, i.e., $\langle u_N/u_{0.4D} \rangle_{1\,m/s}$, at varying incoming flow speeds, where $\langle\ \rangle_{1\,m/s}$ denotes the ensemble average over a bin of 1 m/s around $u_{0.4D}$. Note that the bins with less than 10 s data are excluded in this analysis to ensure statistically significant results. As shown in the figure, $\langle u_N/u_{0.4D} \rangle_{1\,m/s}$ decreases first with increasing incoming wind speed (characterized by $u_{0.4D}$) up to around 11 m/s ($U_{\text{rated}}$), and then plateaus, and finally rises after the wind speed exceeds around 14 m/s. Such trend is likely to be caused by two competing effects, i.e. the induction effect to decrease $\langle u_N/u_{0.4D} \rangle_{1\,m/s}$ and the effect of the flow acceleration around the nacelle to increase the ratio. Specifically, the sonic-SLPIV ratio starts at around 0.98 at the lower wind speeds (~8 m/s) during the period of our deployment. Within this speed range, the turbine operates at region 2.5 with the pitch angle ($\beta$) changing from 1° – 4°. The value of sonic-SLPIV ratio indicates that the effect of induction and nacelle acceleration almost cancel each other at around 8 m/s. As the incoming flow speed increases, the turbine increases its power extraction, accompanied by a stronger induction effect and thus a decreased ratio. As incoming flow speed continue to increase to be larger than $U_\text{rated}$, the turbine enters region 3, where $\beta$ varies significantly above 4° to mitigate the adverse effects of wind load on the turbine structure and the power extraction stays unchanged. Considering the effect of flow acceleration around the nacelle increases with incoming flow, the combined effects result in a plateau of the ratio. As the incoming wind speed continues to ramp up, the flow acceleration effect becomes dominant, causing the ratio to increase.

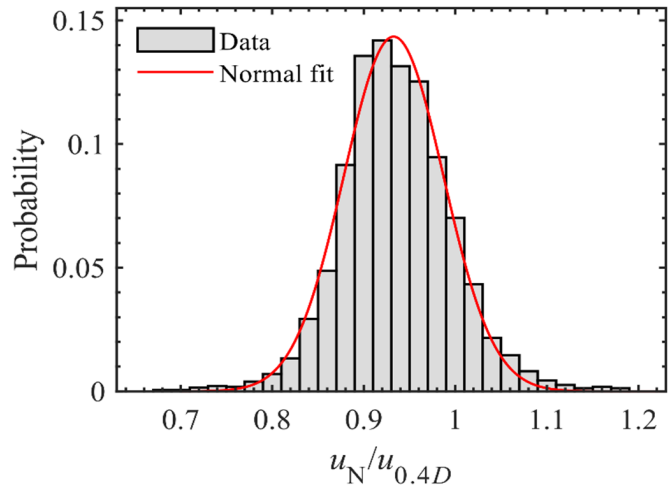

Figure 11. The histogram of the sonic-SLPIV velocity ratio in the period of our field deployment.

To show the statistical distribution of the mismatch between the incoming flow and the nacelle measurements, Figure 10 provides the histogram of the sonic-SLPIV velocity ratio with a bin size of 0.02. The distribution matches well to a normal distribution with a mean of 0.94 and a standard deviation of 0.06. Moreover, 85% of the data yield a ratio smaller than 1, indicating induction plays a more dominant role compared to the effect of flow acceleration around the nacelle. Similar to the variation of NTFs mentioned in the literature, the spread of the sonic-SLPIV velocity ratio observed here could be caused by the varying wind turbine and atmospheric conditions [36,39,41]. It is worth noting that according to the literature [31,48,49], the turbine yaw error ($\Delta\gamma$) also follows a normal distribution. As the turbine yaw has been shown to affect both the power extraction [49]and the unsteady flows around nacelle [41], it is thus suggested that the statistical distributions of $u_N/u_{0.4D}$ and $\Delta\gamma$ are closely related.

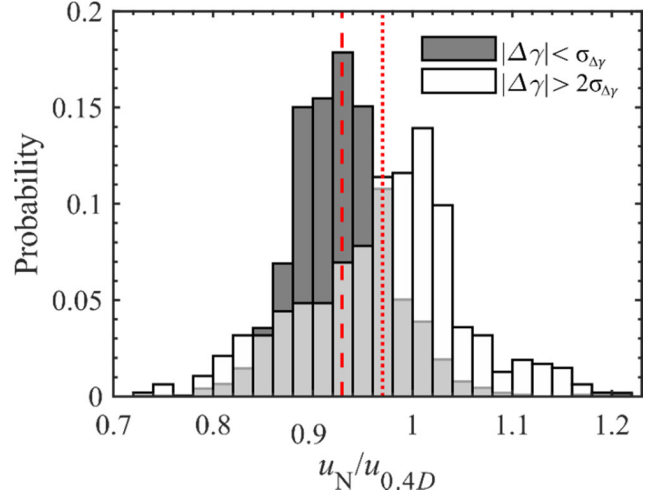

Figure 12. The histograms of sonic-SLPIV velocity ratio under low yaw errors (i.e. $|\Delta\gamma| < \sigma_{\Delta\gamma}$) and high yaw errors (i.e. $|\Delta\gamma| > 2\sigma_{\Delta\gamma}$).

To further explore the connection between turbine misalignment and nacelle measurement accuracy, Figure 11 presents the histograms of sonic-SLPIV ratio corresponding to low and high yaw errors. Note that the yaw error in our data also satisfies the normal distribution consistent with the literature mentioned above, and has a zero mean and standard deviation $\sigma_{\Delta\gamma} = 5°$. The low and high yaw errors are defined using $|\Delta\gamma| < \sigma_{\Delta\gamma}$ and $|\Delta\gamma| > 2\sigma_{\Delta\gamma}$, respectively. For $|\Delta\gamma| < \sigma_{\Delta\gamma}$, the distribution of $u_N/u_{0.4D}$ stays approximately normally distributed with a mean value of 0.93 and a slight skewness towards lower values of $u_N/u_{0.4D}$. In comparison, for $|\Delta\gamma| > 2\sigma_{\Delta\gamma}$, the distribution deviates largely from the normal distribution and spreads over a wider range from 0.7 to 1.2 with the mean shifting to a larger value of 0.97. This trend suggests that larger yaw error induces stronger wind fluctuation and statistically enhances the effect of flow acceleration around the nacelle, consistent with previous numerical simulations of the flow around the nacelle [41]. Besides the statistical evidence shown in Figure 11, the correlation between $u_N/u_{0.4D}$ and $\Delta\gamma$ can be also observed from the time series of these signals under selected occasions. Figure 12 presents a sample time series of $u_N/u_{0.4D}$ and $|\Delta\gamma|$ in a period of 250 s. The figure clearly illustrates a direct correspondence between these two signals, i.e., the rising of $|\Delta\gamma|$ is accompanied with the increase of $u_N/u_{0.4D}$. Note that such clear correspondence over long time duration is not commonly observed in the period of our deployment, indicating that other factors play a role in the variation of $u_N/u_{0.4D}$.

To explore the effect of other factors on the nacelle sonic measurements, Figure 13 presents the ensemble-averaged sonic-SLPIV velocity ratio under different incoming flow incident angles ($\alpha$). As shown in Figure 13(a), $\alpha$ is defined as the angle between instantaneous incoming flow velocity and $x$ axis, with counter clock-wise being positive. It is calculated based on SLPIV velocity spatially averaged over a radius of 3 m centered at $x/D \simeq -0.4$. To exclude the effect of yaw error on the analysis, data with larger yaw error (i.e., $|\Delta\gamma| > \sigma_{\Delta\gamma}$) has been filtered out leaving 57.1 % of data points from all runs. The corresponding ensemble averaged velocity ratio is calculated using a bin width of $\alpha = 2°$ and denoted as $\langle u_N/u_{0.4D}\rangle_{2°}$. Figure 13(b) shows $\langle u_N/u_{0.4D}\rangle_{2°}$ for varying $\alpha$ with error bars being the root-mean-squared (rms) value for each bin. For $\alpha < 0°$, the ratio decreases as the angle approaches 0°, with the rms value remaining at a constant level of around 0.05. This trend can be explained by the shift of wind speed sampling elevation depending on the incident angle. Specifically, with downward flow ($\alpha < 0°$), wind at higher elevations with larger speed are advected to the nacelle sonic causing the ratio to increase. By the same logic, for upward flow ($\alpha > 0°$), we would expect the ratio to decrease with increasing $\alpha$. However, this is not the case as shown in the figure. The ratio plateaus around 0.93 and the rms value rises with increasing $\alpha$. Considering the difference between downward and upward flow relative to the nacelle, this trend may be caused by the nacelle disturbance of the flow. While downward flow is sampled directly after rotor without passing through the nacelle, upward flow interacts with nacelle structures first before sampling by the nacelle sonic. The interaction results in potential acceleration and increased fluctuation of the flow. In other words, even though lower elevation wind with lower speed is advected to the nacelle location, due to its interaction with the nacelle, the flow is accelerated and becomes increasingly unsteady. Consequently, the ratio remains almost constant and fluctuates at a higher level.

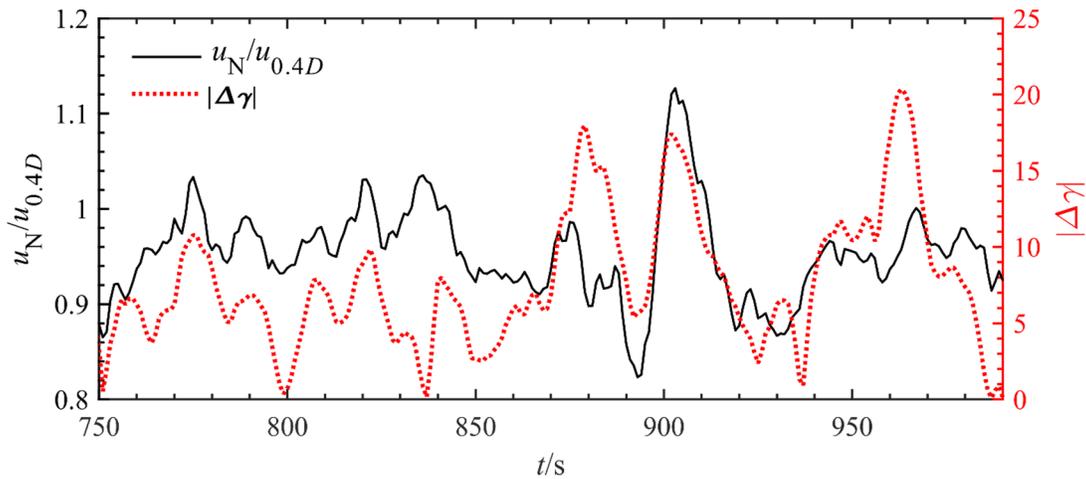

Figure 13. Sample time series of the sonic-SLPIV velocity ratio and the corresponding absolute value of yaw error in the same period.

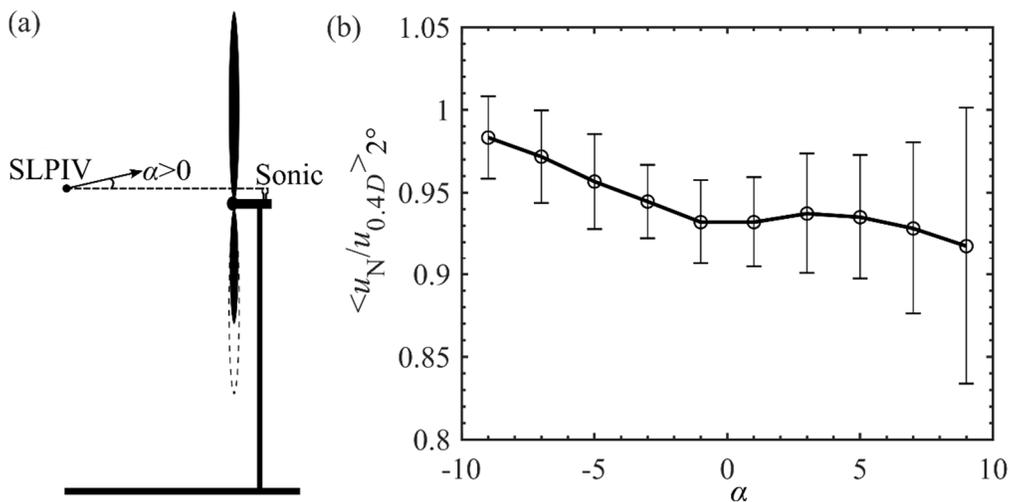

Figure 14. Effect of incoming flow incident angle on the nacelle sonic measurements: (a) schematic showing the effect of incoming flow incident angle ($\alpha$), and (b) ensemble-averaged sonic-SLPIV velocity ratio with bin size of 2° for $\alpha$.

An investigation of the effect of velocity fluctuation on the sonic-SLPIV velocity ratio is also conducted as shown in Figure 14. The intensity of velocity fluctuation is characterized using the ratio between rms and mean incoming flow velocity calculated over a specified time period, i.e. $u_{0.4D,rms}/U_{0.4D}$. To capture the variation of the intensity of velocity fluctuation across our deployment with both sufficient temporal resolution and statistical robustness, a window of 20 s is selected. Sensitivity analysis on this choice has been conducted with varying window size ranging from 10 s to 60 s resulting in a similar trend. As in the previous investigation on incident angle effect, data with $|\Delta\gamma| > \sigma_{\Delta\gamma}$ has also been excluded. Figure 14 presents the corresponding ensemble averaged sonic-SLPIV velocity ratio calculated using a bin size of 0.02 for $u_{0.4D,rms}/U_{0.4D}$, denoted as $\langle u_N/u_{0.4D}\rangle_{0.02}$ and the error bars represent the rms values for each bin. The ratio shows little correlation with velocity fluctuation intensity. Such observations can be explained with the scattered information from the literature. Specifically, Martin et al. [39] showed increasing turbulence intensity can cause enhanced induction effect within a certain range of wind speeds On the other hand, field measurements suggests an increase in standard deviation of yaw error correlates positively with increasing turbulence level [31,48,49]. According to our discussion for Figure 11, increase of yaw error fluctuation may lead to enhanced flow acceleration around the nacelle. Accordingly, our results here suggest that the two competing effects that determine the accuracy of nacelle sonic measurement may be equally influenced by velocity fluctuation at the time scale used in our analysis, leading to the reduction of the influence of velocity fluctuation on the ratio overall.

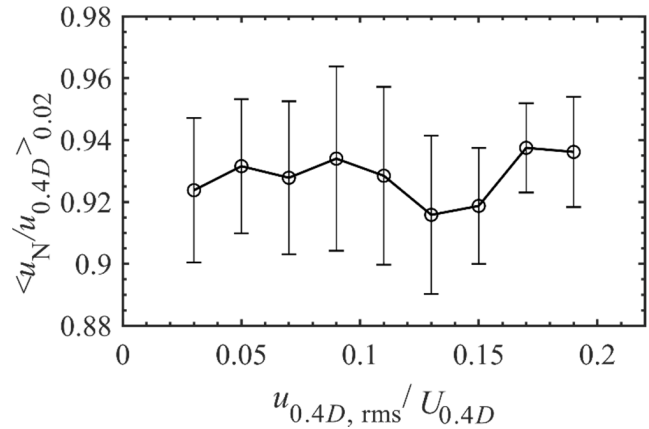

Figure 15. Effect of velocity fluctuation on the nacelle sonic measurements. The figure shows ensemble-averaged sonic-SLPIV velocity ratio with bin size of 0.02 and the velocity fluctuation intensity estimated using a window size of 20 s.

## 4. Summary and Discussion

The current study reports incoming flow measurements of a 2.5 MW pitch-regulated turbine at high spatio-temporal resolution, using super-large-scale particle image velocimetry (SLPIV) with natural snowflakes. The datasets include over one hour duration of incoming flow with an effective field of view 85 m (vertical) × 40 m (streamwise) centered at $0.2D$ upstream. Consistent with previous studies, the mean flow field shows the presence of the induction zone where mean streamwise velocity decreases approaching the rotor plane for all elevations and a distinct region with enhanced vertical velocity. In comparison to predictions from vortex theory, SLPIV streamwise velocity at hub height shows a steeper velocity drop close to the rotor plane and a more confined induction effect with the velocity drop becoming negligible at $x/D \simeq -0.4$. Accordingly, using the SLPIV data at $x/D \simeq -0.4$ as a surrogate of the undisturbed incoming flow, we also investigated the accuracy of nacelle-mounted sonic measurements through comparison with SLPIV at the same elevation where the two measurements are most correlated.

Time series of nacelle sonic and SLPIV measured streamwise velocity show generally matched trends with time-varying discrepancies potentially due to the induction effect and the flow acceleration around the nacelle. These discrepancies between the two signals, characterized by the sonic-SLPIV velocity ratio, is normally distributed with 85% of the time less than unity. The velocity ratio first decreases with increasing incoming wind up to around the rated speed of the turbine, then plateaus, and finally rises with further increase of wind speed. With conditional sampling, the statistical distribution of the velocity ratio shows that larger yaw error leads to an increase in both the mean and the spread of the distribution which may be associated with enhanced flow acceleration around the nacelle. Moreover, as the incident angle of the incoming flow changes from negative to positive (i.e. from pointing downward to upward), the velocity ratio first decreases as the angle approaches zero. With further increase of the incidence, the ratio then plateaus with augmented fluctuation. Finally, our results show that the intensity of short-term velocity fluctuation has limited impact on the sonic-SLPIV velocity ratio.

Our study provides a characterization of the incoming flow in the near field of a utility-scale turbine at unprecedented spatio-temporal resolution using the SLPIV technique. The result can serve as a valuable benchmark dataset for validating high-fidelity numerical simulation and potentially improving the existing incoming flow models derived from prior studies. Using this dataset, our study conducted a detailed assessment of the performance of the nacelle-mounted sonic anemometer as the only incoming flow sensor for the majority of utility-scale turbines nowadays. Such assessment provides useful information for physical understanding of the variation of nacelle transfer function under different turbine conditions. More importantly, this information may shed light on better utilization of nacelle sonic measurements or further improvement of incoming flow monitoring for turbine controls.

Needless to say, our current study still has a number of limitations. First, our measurements are limited to the region of less than half of the rotor diameter upstream the turbine, and to a lower half of the rotor plane with very few prior studies available for direct comparison. The measurements were only conducted at one spanwise location with respect to the turbine. Particularly, we acknowledge that the SLPIV data at $x/D \simeq -0.4$ as a surrogate of the undisturbed incoming flow may still be subject to a slight induction effect. However, the primary reason for us to choose the SLPIV closer to the turbine is to explore the correlation of instantaneous signals between the incoming flow and nacelle sonic as opposed to the long-term averages presented in the literature. The instantaneous measurements made further from the rotor plane results in quick decorrelation with the wind field ultimately interacting with the rotor as pointed out by [31]. Second, the present data has a limited time duration and span of wind and turbine operational conditions. This limitation precludes us from effectively using conditional sampling to fully assess the impact of different physical parameters on the nacelle measurement accuracy. To address these limitations, future field deployments will be conducted at multiple streamwise and spanwise locations upstream of the turbine and under different field conditions. Depending on the snow conditions, our measurement also has the potential to extend our sampling area to capture the flow field over full vertical span of the rotor. In the end, it is worth noting that the analysis made in the paper relies only on a traditional nacelle anemometer setup after the rotor plane and does not depend on any additional of geometric information of the system. Although instantaneous nacelle sonic measurements may depend on the specific configuration of sonic sensors on the nacelle as suggested in a recent numerical study [41], the physical understanding and the overall trends derived from our study can be generally applicable to other utility-scale turbine cases.

# Acknowledgement

This work was supported by the National Science Foundation CAREER award (NSF-CBET-1454259), Xcel Energy through the Renewable Development Fund (grant RD4-13) as well as IonE of University of Minnesota. We also thank the students and the engineers from St Anthony Falls Laboratory, including S. Riley, T. Dasari, Y. Wu, J. Tucker, C. Ellis, J. Marr, C. Milliren and D. Christopher for their assistance in the experiments.